\pdfoutput=1
\documentclass[fleqn,usenatbib]{mnras}

\usepackage{newtxtext,newtxmath}

\usepackage[T1]{fontenc}

\usepackage{graphicx}
\usepackage{amsmath}

\usepackage{color}

\title[Aeolian Erosion in WD Disks]{Rapid destruction of planetary debris around white dwarfs through aeolian erosion}

\author[Rozner et al.]{
Mor Rozner,$^{1}$\thanks{E-mail: morozner@campus.technion.ac.il}
Dimitri Veras,$^{2,3}$\thanks{STFC Ernest Rutherford Fellow}
Hagai B. Perets$^{1}$
\\
$^{1}$ Physics department, Technion - Israel Institute of Technology, Haifa, 320004, Israel\\
$^{2}$Centre for Exoplanets and Habitability, University of Warwick, Coventry CV4 7AL, UK\\
$^{3}$Department of Physics, University of Warwick, Coventry CV4 7AL, UK
}

\date{Accepted XXX. Received YYY; in original form ZZZ}

\pubyear{2020}

\begin{document}
\label{firstpage}
\pagerange{\pageref{firstpage}--\pageref{lastpage}}
\maketitle

\begin{abstract}
The discovery of numerous debris disks around white dwarfs (WDs), gave rise to extensive study of such disks and their role in polluting WDs, but the formation and evolution of these disks is not yet well understood. Here we study the role of aeolian (wind) erosion in the evolution of solids in WD debris disks.  Aeolian erosion is a destructive process that plays a key role in shaping the properties and size-distribution of planetesimals, boulders and pebbles in gaseous protoplanetary disks. Our analysis of aeolian erosion in WD debris disks shows it can also play an important role in these environments. We study the effects of aeolian erosion under different conditions of the disk, and its erosive effect on planetesimals and boulders of different sizes. We find that solid bodies smaller than $\sim 5 \rm{km}$ will be eroded within the short disk lifetime. We compare the role of aeolian erosion in respect to other   destructive processes such as collisional fragmentation and thermal ablation. We find that aeolian erosion is the dominant destructive process for objects with radius $\lesssim 10^3 \rm{cm}$ and at distances $\lesssim 0.6 R_\odot$ from the WD. Thereby, aeolian erosion constitutes the main destructive pathway linking fragmentational collisions operating on large objects with sublimation of the smallest objects and Poynting-Robertson drag, which leads to the accretion of the smallest particles onto the photosphere of WDs, and the production of polluted WDs. 
\end{abstract}


\section{Introduction}
White dwarfs (WDs) are the final evolutionary stage of the vast majority of all stars. Metal-polluted WDs, which represent $25-50 \%$ of the whole WD population \citep{Zuckerman2010,Koester2014}, constitute an observational signature of the accretion of heavy elements. A few per cent of the WDs are surrounded by planetary debris disks \citep{Manser2020}, and over sixty WD disks have already been discovered \citep{Zuckerman2010, Farihi2016}. Most of the observed disks are dusty, although about twenty disks with gaseous components were discovered as well \citep{Dennihy2020,Melis2020,Gentile2020}. 

The quickly growing population of these disks motivates theoretical explorations of their structure and evolution, which can reveal both how they are formed \citep{Veras2014,Veras2015,MalamudPerets2020,MalamusPerets2020II} and how they accrete onto WDs photospheres, creating observable metal pollution \citep{JuraYoung2014,Hollands2018,Harrison2018,Doyle2019,Swan2019,Bonsor2020}.
An interesting observed feature of WD disks is that nearly all the disks are now thought to showcase flux variability on timescales of weeks to decades \citep{Farihi2018,Xu2018,Swan2020}.

The abundance of objects with different sizes in WD disks -- from grain size to minor planets, reveals rich and interesting dynamics (see a detailed review of post-main sequence evolution in \citealp{Veras2016}), which includes many physical processes, among them are replenishment and accretion. Frequent collisions in WD disks \citep{Jura2008,MetzgerRafikovBochkarev2012,KenyonBromley_analytic2017,KenyonBromley2016a,KenyonBromley2017b} lead to replenishment of grains in the disk -- where large objects break into smaller ones, and gradually are removed by radiation pressure and accretion.
The debris disks might give rise to dynamical excitations and perturbations of mass that will eventually drive matter onto the WD within the disk lifetime \citep{Girven2012,VerasHeng2020}. Poynting-Robertson drag results from the radiation force and causes loss of angular-momentum for small pebbles, carrying them to the WD \citep{Burns1979,Rafikov2011,Rafikov2011a}.

The dynamics and architecture of WD disks are similar to protoplanetary disks in some aspects, and different in others. While there are some processes that take place in both of them, such as fragmentation, their parameters might differ significantly.
The scales of WD disks are much smaller, the density and temperature profiles are different, and the typical velocities could be much higher. 

One important process in protoplanetary disks is aeolian (wind) erosion.
Aeolian erosion is a purely mechanical destructive process, which is very common in many occasions in nature, mainly discussed in the context of sand dunes \citep{Bagnold1941}
. Recently, we showed that aeolian erosion can play an important role in planet formation by setting a new growth-barrier for pebbles/boulders in protoplanetary disks, and affecting pebble accretion  and streaming instability \citep{Erosion2,Erosion1}. 
Aeolian erosion in protoplanetary disks is rapid and efficient, as also verified in lab experiments and numerical simulations of the conditions of protoplanetary disks \citep{Paraskov2006,Demirci2020,Demirci2020_2,Demirci2020_3,Schaffer2020}.
Aeolian erosion leaves signatures on the dynamics of objects in the disk, e.g. it fuels pebble-accretion and induces a redistribution of sizes in the disk, and might potentially lead to reshaping or complete destruction of objects \citep{Erosion2,Erosion1}. 
In contrast with protoplanetary disks, the temperatures in WD disks are high enough to maintain thermal ablation \citep{Podolak1988,Pollack1996,DangeloPodolak2015} along with aeolian erosion, and accelerate the destruction of small objects.

 In this paper, we suggest that aeolian erosion could take place in WD disks and interact symbiotically with other dynamical processes in the disk. In fact, we show that for certain regimes, for objects with radius $\lesssim 10^3 \rm{cm}$ and at distances $\lesssim 0.6 R_\odot$ from the WD, in a disk with $M_{disk}=10^{24} \rm{g}$ for aeolian erosion is the dominant destruction process. The dominance regime changes with the disk and object's parameters. As we will discuss later, the mass of the disk should be $\gtrsim 10^{21}\rm{g}$ to enable aeolian erosion. 

The efficiency of aeolian erosion, as well as of thermal ablation, depends on the composition of the object and the disk temperature, and acts to grind down large objects into small pebbles. For small pebbles, Poynting-Robertson drag becomes efficient and carries the pebbles onto the WD.  Hence, aeolian erosion might set a lower limit on the accretion flux onto the WD and enhance the disk gas replenishment, by assuring repeatedly regenerating the large abundance of small objects. Moreover, as we showed in \cite{Erosion2}, aeolian erosion induces redistribution of the particle sizes and their abundance in the disk. The size-distribution of objects in WD disks as well as the sizes of objects which accrete on WDs are not well constrained observationally, although some theoretical constraints have been established \citep{KenyonBromley2017a,KenyonBromley2017b}. Aeolian erosion might shed light on in this direction as well. 

The paper is organized as follows: in section \ref{sec:wd disks intro} we briefly review the parameters space and various models of WD disks. In section \ref{sec: aeolian erosion and thermal intro} we review the models of aeolian erosion and thermal ablation in WDs disks. In section \ref{sec:results} we present our results and include eccentric orbits, multilayer objects and the relationship with thermal effects. In section \ref{sec: discssion and implications} we discuss our results and suggest possible implications: we discuss the dependence of our model on the disk parameters, the symbiotic relations with other processes in the disk including collisional cascade, external seeding and further disk generations. In section \ref{sec:caveats}, we discuss the caveats and limitations of our study. In section \ref{sec:summary} we summarize the paper and suggest future directions.   

\section{White Dwarf Disks}\label{sec:wd disks intro}
In contrast with protoplanetary disks which properties are better-constrained, WD disk parameters are uncertain by orders-of-magnitude. Here we briefly review the ranges of these parameters that will be used in the rest of the paper.

The total mass of a WD disk ranges between $10^{12}-10^{25} \rm{g}$ (see a detailed discussion in \citealp{MetzgerRafikovBochkarev2012,VerasHeng2020} and references therein). The mass of the gas in the disk, parametrized by the gas-to-dust ratio, is highly unconstrained and ranges between $10^{-5}$ and unity \citep{Veras2016}. Observations set the lower limit of the inner-radius to be $\lesssim 0.2 R_\odot$ (e.g. \citealp{Rafikov2011}) and the outer-radius to be $\gtrsim 1.2 R_\odot$ (e.g. \citealp{Gansicke2006}).

The surface density profile is given as \citep{MetzgerRafikovBochkarev2012},

\begin{align}\label{eq: surface density}
\Sigma_g (a) = \Sigma_{g,0} \left(\frac{4.72a}{0.6 R_{\odot}}\right)^{-\beta}\left(\frac{M_{{\rm disk}}}{10^{24} \rm{g}}\right)
,
\end{align}

\noindent

 where $a$ is the distance from the center of the disk, 
 $\Sigma_{g,0}$ is the fiducial surface density and $\beta$ is an arbitrary exponent which is parametrized as $\beta = n+1/2$, where $n$ describes the viscosity power law  $\nu(a)\propto a^n$. The gas density $\rho_g$ is determined by the surface density, 

\begin{align}
\rho_g = \frac{\Sigma_g}{2 h(a)}
,
\end{align}

\noindent
where $h$ is the height of the disk. 

This scaling of $\beta$ also induces a temperature scaling \citep{MetzgerRafikovBochkarev2012}, 

\begin{align}
T(a)\propto \begin{cases}
{\rm constant}, \ n = 3/2 ; \\
a^{-1/2}, \ n=1  
\end{cases}
\end{align}

\noindent
where $n=3/2$ corresponds to an optically thick disk, and $n=1$ corresponds to an optically thin disk.
The aspect ratio of the disk is loosely constrained and strays over some orders of magnitude. We will use an aspect ratio of $10^{-2}$ unless stated otherwise. 
\noindent
 The aforementioned choice of parameters yields typical values of $\Sigma_g= 5.1\times 10^3 g \ cm^3, \ \rho_g=2.55\times 10^{-5} \rm{g \ cm^{-3}},$ and $T = 10^3 \rm{K}$ at $a=0.6 R_\odot$. 
 
Unless stated otherwise, we will use $n=3/2$ for all our profiles. 

The lifetime of WD disks (at least the ones within a couple of Solar radii from the WD) ranges between $\sim 10^4-10^6 \  \rm{yr}$ \citep{Girven2012,VerasHeng2020}.
\noindent
\cite{Girven2012} estimated the age of WD disks assuming a constant accretion rate to the WD. Recently, \cite{VerasHeng2020} introduced a different estimation method, arising from the dynamical processes that the disk should go through and their typical timescales.
\noindent

\section{Aeolian erosion \& Thermal Destructive Processes in Disks}\label{sec: aeolian erosion and thermal intro}

\subsection{Aeolian erosion}

Consider a spherical object with radius $R$ that resides at a constant distance $a$ from the WD and is built from grains with a typical size $d$, that move in a gaseous medium with a density $\rho_g$.  The pressure support present in gaseous disks leads to a difference between the Keplerian velocity $\Omega_K$ around the WD and the actual angular velocity $\Omega_g$,  $\Omega_g-\Omega_k\approx (2\Omega_K a \rho_g)^{-1}\partial P/\partial r $ where $\partial P/\partial r$ is the pressure gradient. Furthermore, the pressure support leads to the radial drift, which presents one of the fundamental problems in planet formation in protoplanetary disks --  the meter-size barrier \citep{Weidenschilling1977}.

 The radial velocity in protoplanetary disks obtains a maximum for $\sim 1 \rm{m}$ size objects, and these objects inspiral to the inner parts of the disk on timescales shorter than the expected growth timescale. The drift velocity is lower for smaller objects -- they are better coupled to the gas and hence have slower relative velocities; larger objects are loosely coupled to the gas and experience slower drift velocities. For objects in WD disks, the maximal drift velocity is obtained for smaller objects, in size $\lesssim 1 \rm{cm}$  \citep{KenyonBromley2017b}.

\begin{align}
    v_{\rm rel,r}&=-\frac{2\eta v_k \rm St}{1+\rm{St}^2},
    \\ v_{\rm rel,\phi}&= -\eta v_k \left(\frac{1}{1+\rm{St}^2}-1\right) \label{eq:vrel}
\end{align}
 where the Stokes number is defined by

\begin{align}
    {\rm St}= \Omega_K t_{\rm stop},
    \\  t_{\rm stop}= \frac{mv_{\rm rel}}{F_{D}} \label{eq:st}
\end{align}

\noindent

 where $m$ is the mass of the object, $F_D$ is the drag force, which is given by 
\begin{align}
\textbf{F}_D= \frac{1}{2}C_D({\rm Re}) \pi R^2 \rho_g v_{\rm rel}^2 \hat{\textbf{v}}_{\rm rel}    
\end{align}

For the drag coefficient, we adopted an emperical fitted formula, based on experminal data in the regime $10^{-3}\leq Re\leq 10^5$ (\citealp{PeretsMurrayClay2011} and references therein)

\begin{align}
    C_D({\rm Re}) = \frac{24}{{\rm Re}}(1+0.27{\rm Re})^{0.43}+0.47\left[1-\exp\left(-0.04{\rm Re}^{0.38}\right)\right]\label{eq:cd}
\end{align}

\noindent
where $Re$ is the Reynolds number, defined by 

\begin{align}
{\rm Re} = \frac{4R v_{{\rm rel}}}{v_{{\rm th}}\lambda} 
,
\end{align}

\noindent
where $v_{\rm{th}} = (8/\pi)^{1/2}c_s$ is the mean thermal velocity, $c_s = \sqrt{k_B T/\mu}$ is the speed of sound, $k_B$ is Boltzmann constant, $T$ is the temperature of the disk, $\mu$ is the mean molecular weight, taken to be $2.3m_H$ (following \citealp{PeretsMurrayClay2011}) with $m_H$ the mass of a hydrogen atom and $\lambda = \mu/(\rho_g \sigma)$ is the mean free path of the gas. 

\begin{figure}
    \includegraphics[width=1.\linewidth]{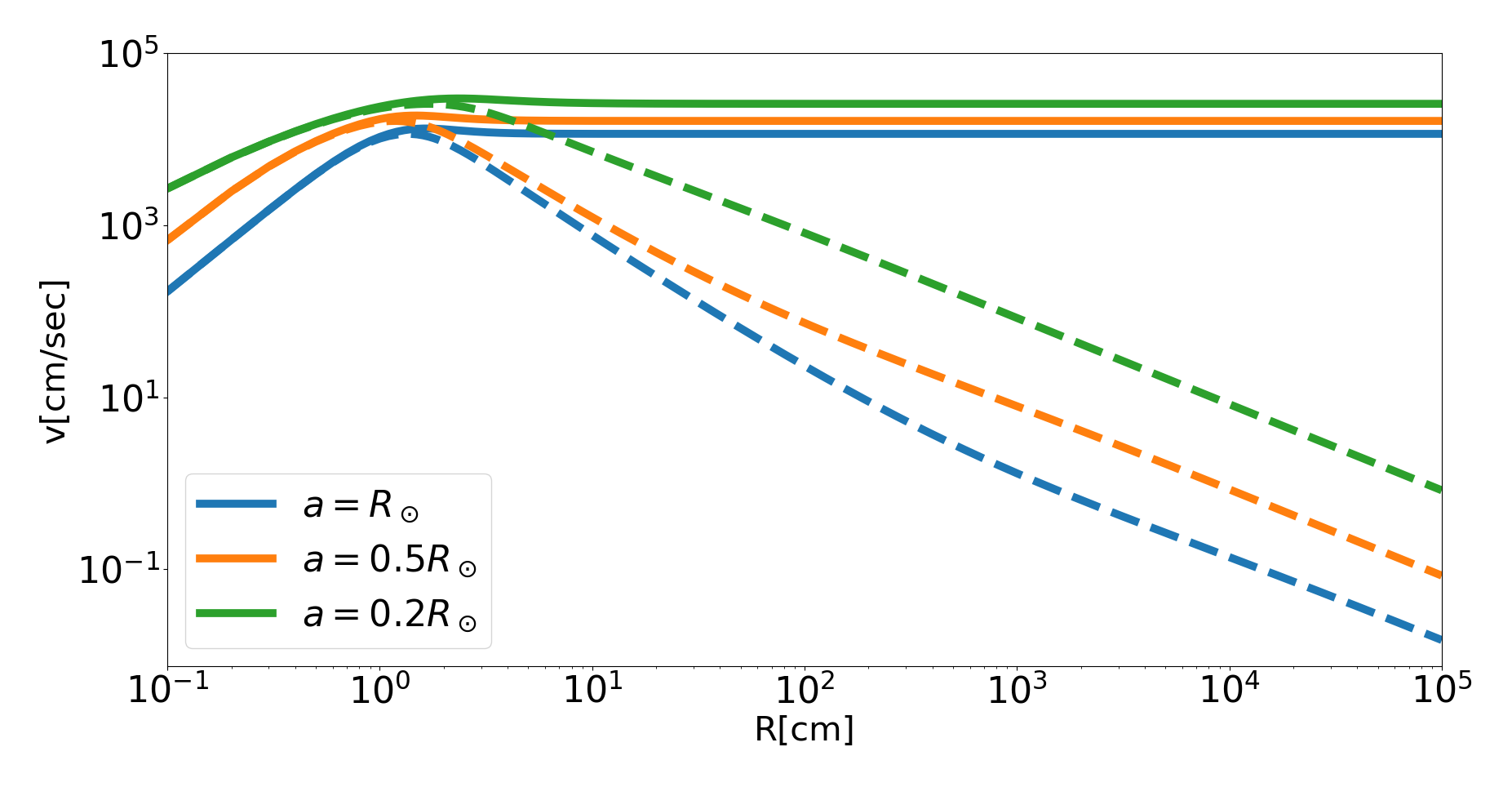}
\caption{The relative velocity between objects and the gas, in different constant distances from the WD. The solid lines are for $v_{\rm rel}$ and the dashed ones are for the radial component, $v_{\rm r}$.
} 
\label{fig:Weidenschilling}
\end{figure}

In fig. \ref{fig:Weidenschilling}, we present the relative velocities between the gas and objects for different sizes. The radial component of the velocity peaks for $\sim 1\rm{cm}$ size objects, and is about $\sim 10^5 \rm{cm/sec}$. Much smaller objects are well-coupled to the gas, which leads to smaller velocities, while much larger objects are weakly-coupled to the gas and hence are not affected significantly by it. The relative velocity could change for different parameters of the disk, and as we have mentioned, the parameter space is currently wide.

Objects in gaseous disks experience gas drag, which depends strongly on their velocity relative to the gas, $v_{\rm{rel}}$. The gas-drag can play the role that is usually played by the wind in aeolian erosion and trigger loss of the outer layer of objects, as long as they reach the threshold conditions. 

The threshold velocity is dictated by the balance between cohesion forces, self-gravity and gas-drag, i.e. the headwind should be stronger than the attraction force between grains in order to initiate aeolian erosion.

The threshold velocity is given by 

\begin{align}
    v_{\star}= \sqrt{\frac{A_N}{\rho_g}\left(\rho_p g d+  \frac{\gamma }{ d} \right)},
\end{align}

\noindent
where  $A_N=1.23\times 10^{-2}$, and \ $\gamma= 0.165\ \rm{g\ s^{-2}}$ are determined empirically from \cite{ShaoLu2000}. Both of them are intrinsic characteristic of the materials that rise from the cohesion forces that hold the particles together -- mostly electrostatic forces and van der Waals forces. The gravitational acceleration is  $g=Gm/R^2$. 
Above the threshold velocity, i.e. $v_{{\rm rel}}>v_\star$, the aeolian erosion rate is given by \citep{Erosion1}, 

\begin{align}\label{eq:eolian-erosion}
 \frac{dR}{dt} = -\frac{\rho_g v_{\rm rel}^3}{4\pi R \rho_p a_{\rm coh}}
\end{align}

\noindent
where $a_{{\rm coh}}$ is the cohesion acceleration. The derivation rises from the work done on the eroded object by the shear pressure; see a detailed derivation from equations 8 and 9 of \cite{Erosion1}.
The derivation is based on estimation of the typical sweeping rate of grains from the outer layer and calculating the work done by them. 

\subsection{Thermal Destructive Processes}\label{subsec:thermal destructive}

Along with aeolian erosion, the high temperatures that are usually found in WD disks might give rise to destructive thermal processes such as thermal ablation  \citep{Podolak1988,Pollack1996,DangeloPodolak2015} and sublimation (e.g \citealp{MetzgerRafikovBochkarev2012,Shestakova2019}). The heat that the outer layer absorbs might lead to phase transitions and then mass loss. There are two regimes, separated by a critical temperature $T_{cr}$, in which the latent heat required for vaporization is zero, and varies according to the material (e.g. \citealp{Opik1958,Podolak1988,DangeloPodolak2015}). See the fiducial parameters for ablation in table \ref{table:parametres_table}. 

Below the critical temperature, the rate in which vaporization removes mass is dictated by Hertz-Knudsen-Langmuir equation; see a discussion in \cite{DangeloPodolak2015}. In this regime, aeolian erosion significantly dominates for our choice of parameters, and the timescale for thermal ablation below the critical temperature is $\gtrsim 95 \ \rm{yrs}$ -- much longer than typical aeolian erosion timescale, which enables us to neglect the thermal effect and focus on the mechanical processes. 

Above the critical temperature, the contribution from thermal processes might add a significant contribution to aeolian erosion and should be added to eq. \ref{eq:eolian-erosion}; assuming blackbody emission, the thermal ablation term is given by

\begin{align}
\frac{dR}{dt}\bigg|_{{\rm ablation}} = \frac{1}{L_s \rho_p}\epsilon_s \sigma_{SB}(T_{cr}^4-T_g^4)
,
\end{align}

\noindent
where $\epsilon_s$ is the thermal emissivity of the object ($\epsilon_s=1$ for a perfect blackbody), $L_s$ is the particle specific vaporization energy, $T_{cr}$ is the critical temperature  -- which depends on composition of the material -- and $T_g$ is the gas temperature. At high enough temperatures, close to the WD, small grains sublimate to gas (e.g \citealp{MetzgerRafikovBochkarev2012,Shestakova2019}).

\section{Results}\label{sec:results}

In this section we present the evolution of objects in WD disks due to aeolian erosion and thermal ablation, starting from a fiducial set of parameters and then vary them.

\begin{figure*}
   \includegraphics[ ,width=.5\linewidth,height=5cm]{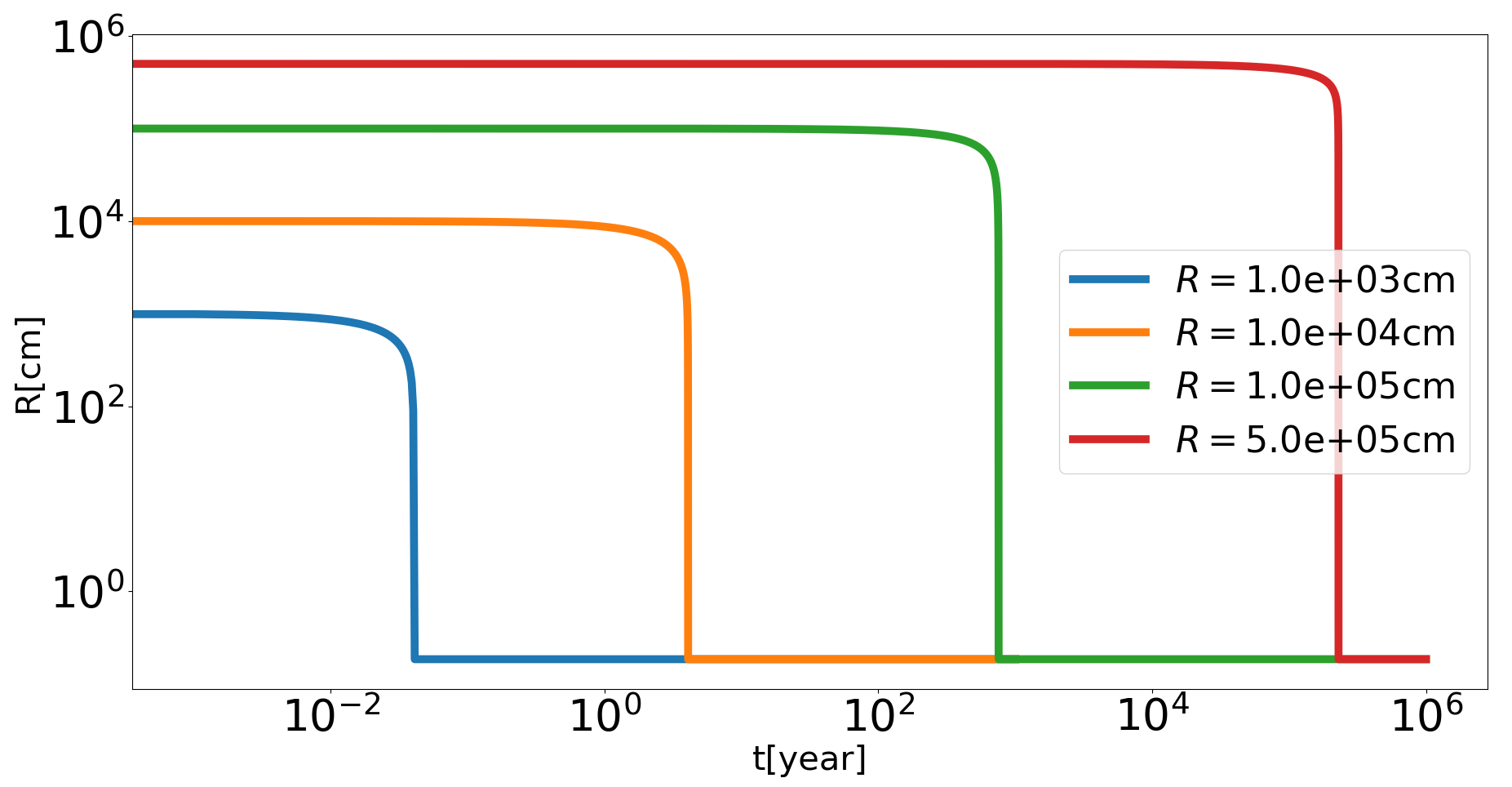}\includegraphics[ width=.5\linewidth,height=5cm]{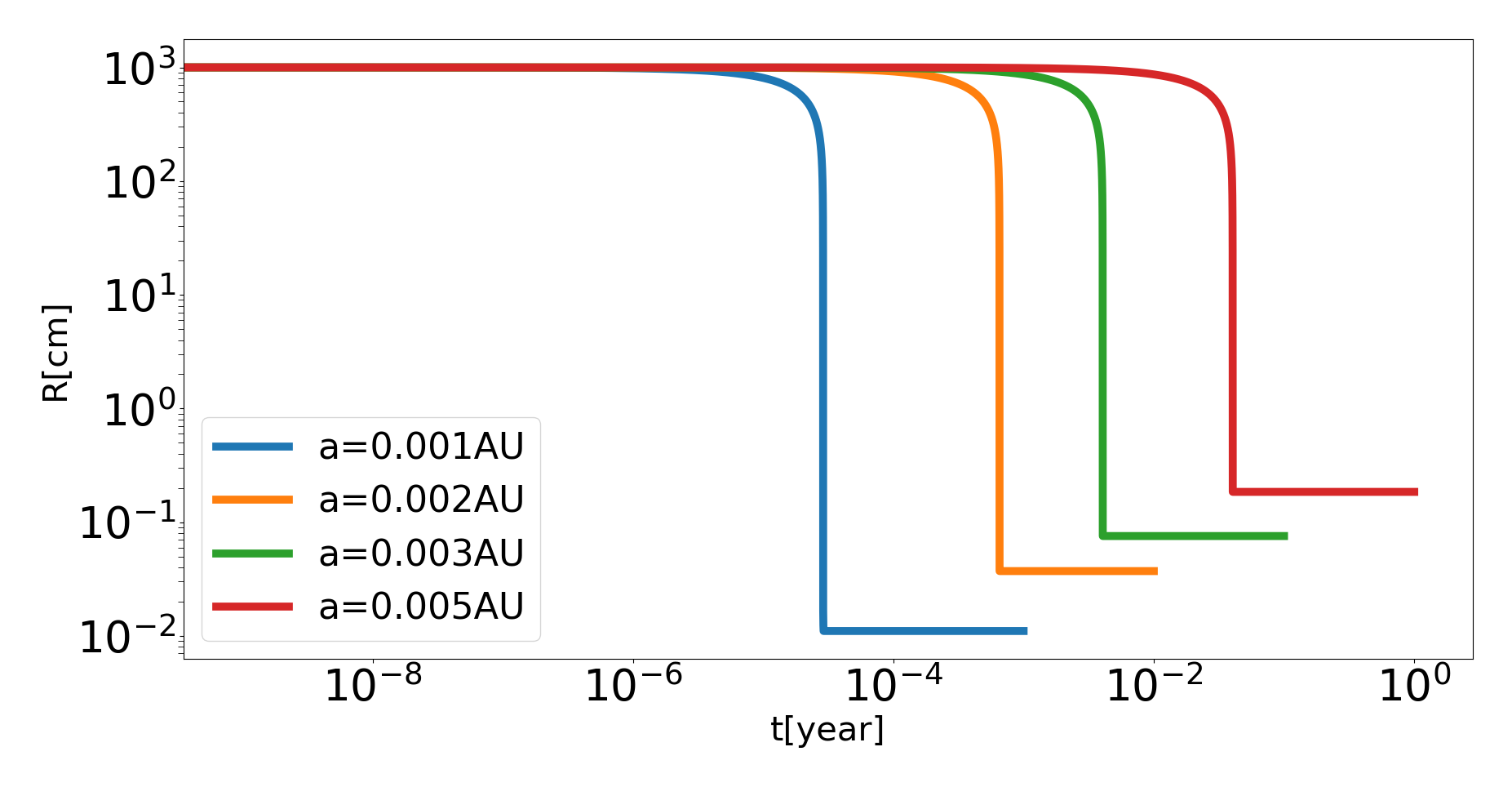}
\caption{The time evolution of objects embedded in a WD disk due to aeolian erosion. {\it Left}: Evolution at a constant distance $a_0=0.005 \rm{AU}\approx 1R_\odot$ from the WD. {\it Right}: Different distances from the WD, with a constant initial size of embedded objects $R_0 = 10^3 \rm{cm}$. }
\label{fig:dynamical}
\end{figure*}

Aeolian erosion in WD disks is quite efficient, as manifested in Fig. \ref{fig:dynamical}. The timescales for aeolian erosion are extremely short, and for a constant distance $a=0.005 \rm{AU}\approx 1 R_\odot$ from the WD disk, objects with radii as large as $\sim 5\times 10^5 \rm{cm}$ are eroded within the expected disk lifetime. The chosen fixed distance from the WD, along with the rest of the parameters that we don't vary currently, dictates the final size to which objects are ground down -- from $\sim 0.01 \rm{cm}$ for $0.001 \rm{AU}$ to $\sim 0.18 \rm{cm}$ for $0.005 \rm{AU}$. 

The final size is determined by the initial conditions of the eroded object, and induces a size re-distribution according to distance from the WD. The outer parts of the disk are more dilute then the inner ones, and since the rate of aeolian erosion is proportional to the gas density, as can be seen in Eq. \ref{eq:eolian-erosion}, aeolian erosion is more efficient in the inner part of the disks, which leads to shorter timescales and smaller final sizes. Aeolian erosion has a Goldilocks region of sizes in which it attains its maximal efficiency, since the relative velocity, which plays a significant role in this process, varies with the coupling of objects to the gas.

\subsection{Eccentric Orbits}
Aeolian erosion depends strongly on the relative velocity (to the third power), as can be seen from Eq. \ref{eq:eolian-erosion}. Hence, its effects on objects on eccentric orbits may differ substantially from the effects in the circular case,  and lead to stronger more significant erosion due to the higher velocities involved, that might even be supersonic and under some conditions lead to the prompt disruption of objects \citep{Demirci2020_3}.

The velocity of a planetesimal in an eccentric orbit is given by

\begin{align}
\textbf{v}_p =v_k \sqrt{2-\frac{r}{a}} \hat {\textbf{v}}_p=\sqrt{GM \left(\frac{2}{r}-\frac{1}{a}\right)}\hat {\textbf{v}}_p
\end{align}

\noindent
where $M$ is the mass of the WD, $v_k$ is the Keplerian velocity, $r$ is the distance and $a$ is the semimajor axis. 

Assuming a circular disk, the gas moves with a velocity 
\begin{align}
\textbf{v}_g = v_k\sqrt{1-\eta}\hat {\textbf{v}}_g
\end{align}

The magnitude of the relative velocity between the gas and the planetesimal is given by 
\begin{align}
v_{{\rm rel}}= |\hat {\textbf{v}}_p-\hat {\textbf{v}}_g|=\sqrt{v_p^2+v_g^2-2v_pv_g {\hat{\textbf{v}}}_g\cdot {\hat{\textbf{v}}}_p}
.
\end{align}

For simplicity, we will assume that the evolution of objects experiencing aeolian erosion is dominated by the maximal relative velocity in the orbit, when the phase between the gas and the objects is maximal -- $\pi/2$, and also assume that the object is at the pericenter, i.e. $r=r_p=a(1-e)$. See \cite{Mai2020} for more detailed results.

\begin{figure}
    \includegraphics[width=1.\linewidth]{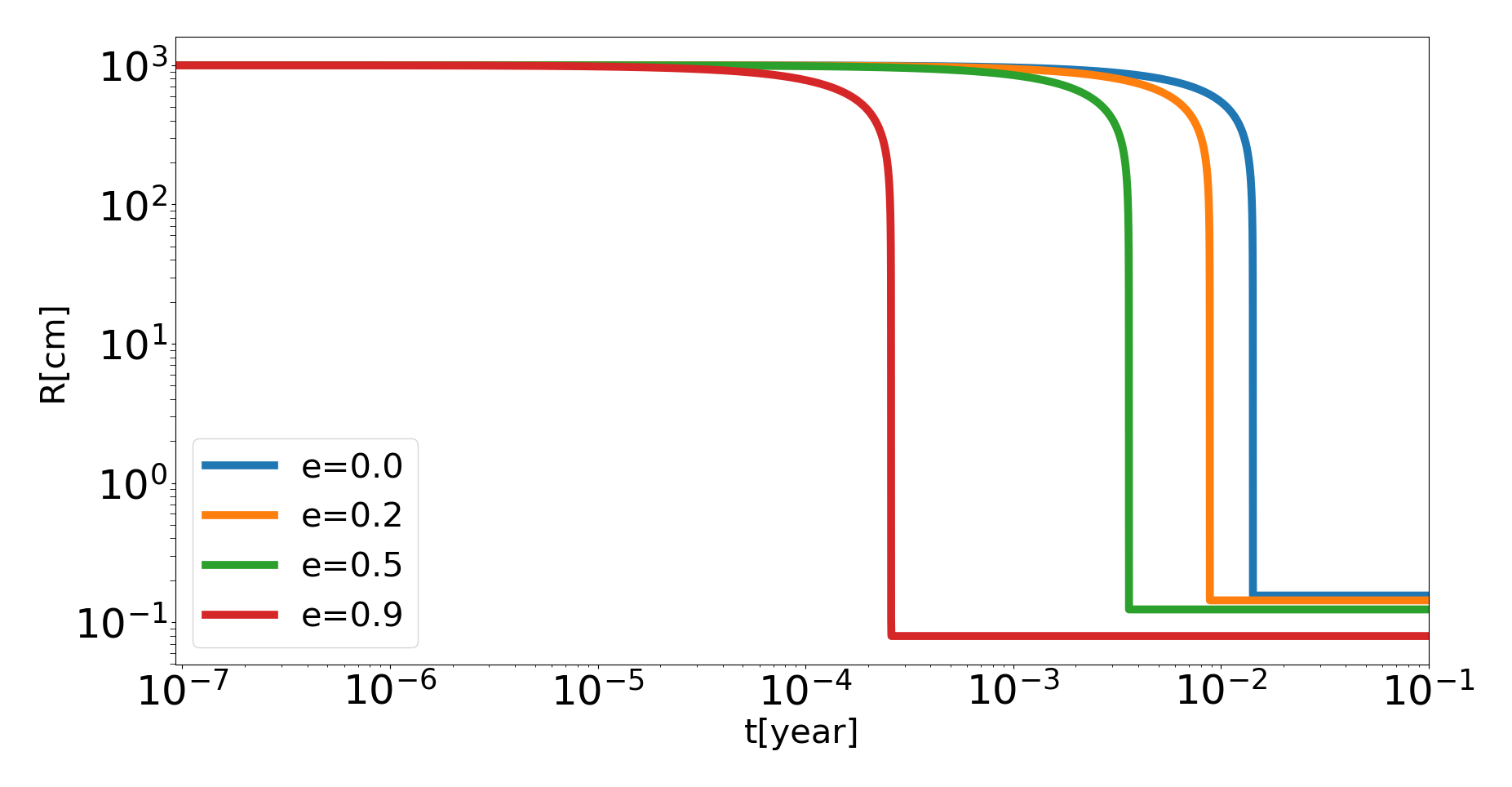}
\caption{The effect of aeolian erosion on an $10^3 \rm{cm}$ object embedded in a WD disk at a distance of $1 R_\odot$ from the center of the WD. The relative velocity of the object is considered as the maximal velocity in the orbit.
} 
\label{fig:eccentric}
\end{figure}

Motion in eccentric orbits shortens the timescales of aeolian erosion and the final size of eroded objects is smaller and could attain $\sim 0.08 \, \rm{cm}$ for $e=0.9$. 

\subsection{Multilayer Objects}
Previously, we discussed objects which are composed from a single-size grain distribution. However, the physical reality might be more complicated and could give rise to a non-trivial internal size distribution. In the following we relax homogeneous composition assumption, and consider the effects of aeolian erosion on inhomogeneous objects. To the best of our knowledge no previous study considered the internal structure of grains within WD disks, and we adopt internal size distribution models usually considered for asteroids.

The Brazil nut effect in asteroids \citep{Matsumura2014} suggests that in a mixture of particles, the larger ones tend to end up on the surface of objects. Therefore, when this process acts, the inner structure of an asteroid is such that the larger grains are in the outer layers.  

\begin{figure}
    \includegraphics[width=1.\linewidth]{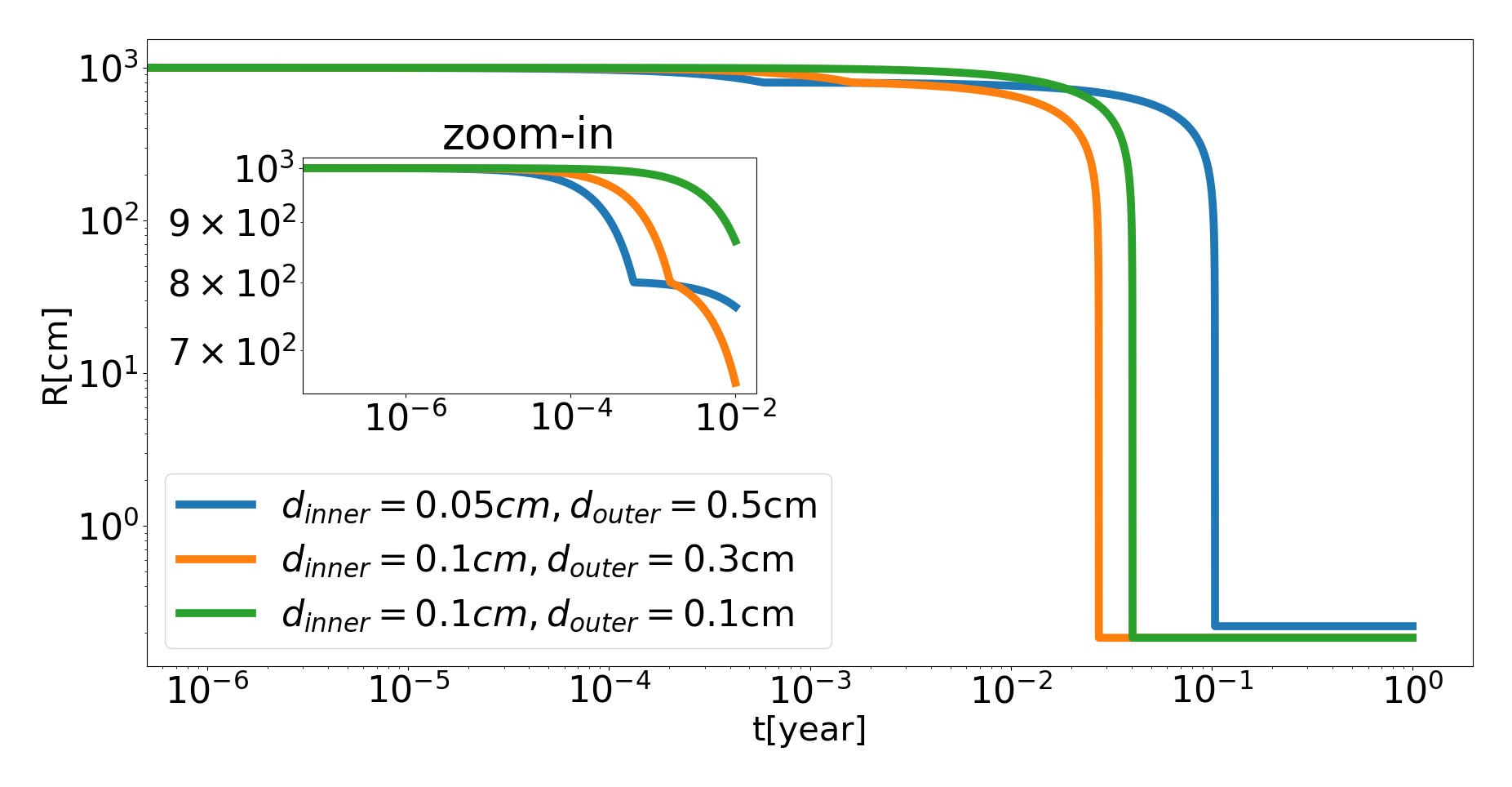}
\caption{The evolution of two-layered differentiated objects (see inset legend), with an outer layer comprising $20 \%$ of the radius and an inner core of the remaining $80\%$ of the radius, under the effects of aeolian erosion; all of the layers have the same density of $3.45 \rm{g \ cm^{-3}}$. The object is embedded in a WD disk and resides at a distance of $0.005 \rm{AU}$ from the host WD. 
} 
\label{fig:differentiated_80_20}
\end{figure}

\begin{figure}
    \includegraphics[width=1.1\linewidth]{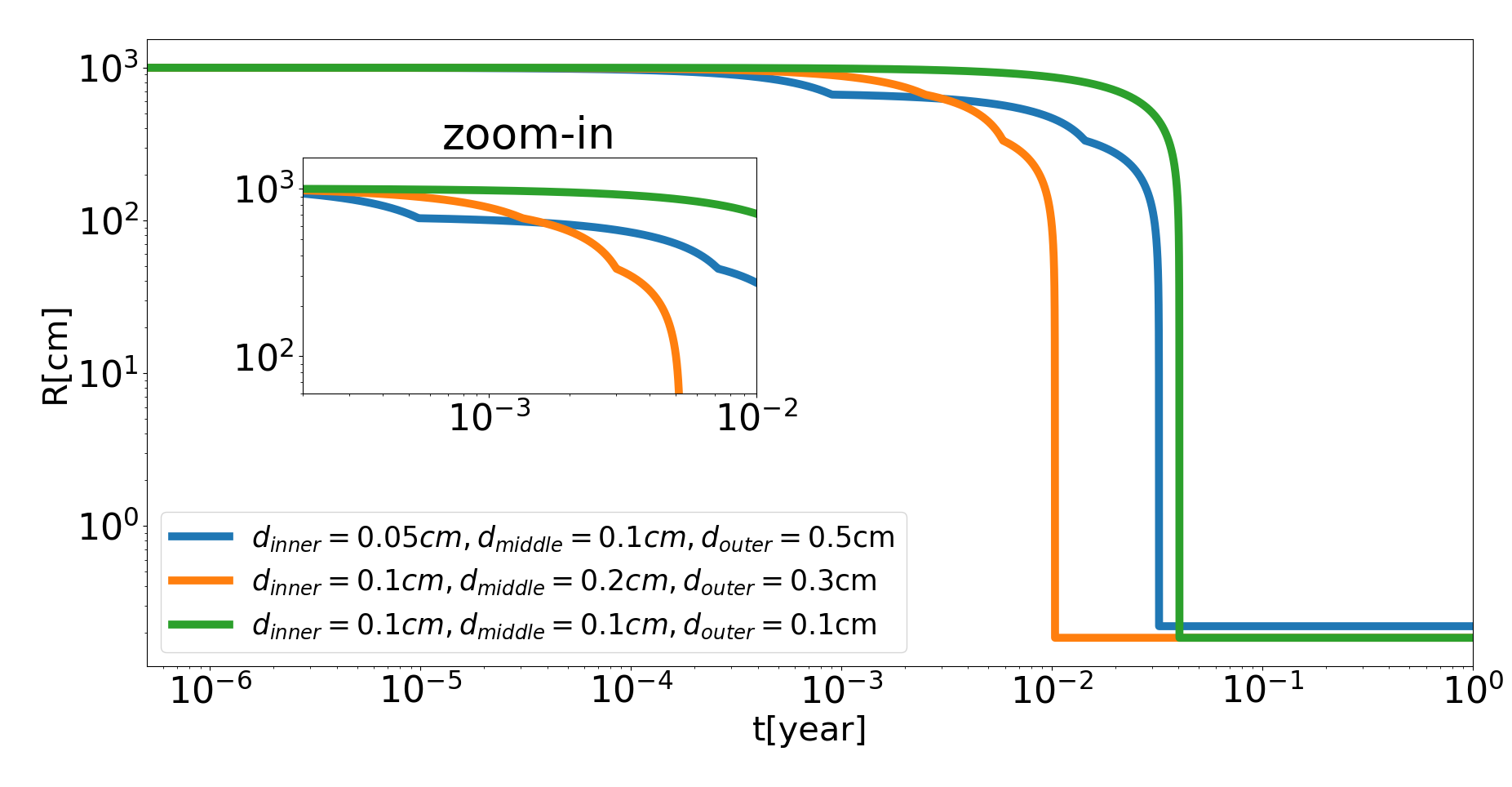}
\caption{The evolution of triple-layered objects, with three equal layers, with each layer comprising $1/3$ of the radius, of the same density $3.45 \rm{g \ cm^{-3}}$, but with different size of inner grains that build each layer (see inset legend). The object is embedded in a WD  disk and resides at a distance of $0.005 \rm{AU}$ from the WD.
} 
\label{fig:differentiated_3_equalayers}
\end{figure}

In Figs. \ref{fig:differentiated_80_20}, \ref{fig:differentiated_3_equalayers}, we present how differentiated objects react to aeolian erosion. Erosion enables us to decompose the layers of an object -- as far as they are in the correct regimes in which erosion is effective -- and to reveal the inner layers in short timescales. Since different sizes of grains impose different rates of aeolian erosion, the total aeolian erosion timescales change such that considering larger grains shortens the timescales, and smaller grains lengthen them.  
Moreover, it can be seen that the final size is determined by the innermost layer, and larger grains lead to smaller final sizes.

\subsection{Thermal Effects}
At high temperatures, thermal processes become more significant and might strengthen the effect of aeolian erosion and give rise to further destruction and shorten the timescales. 

\begin{figure}
    \includegraphics[width=0.99\linewidth]{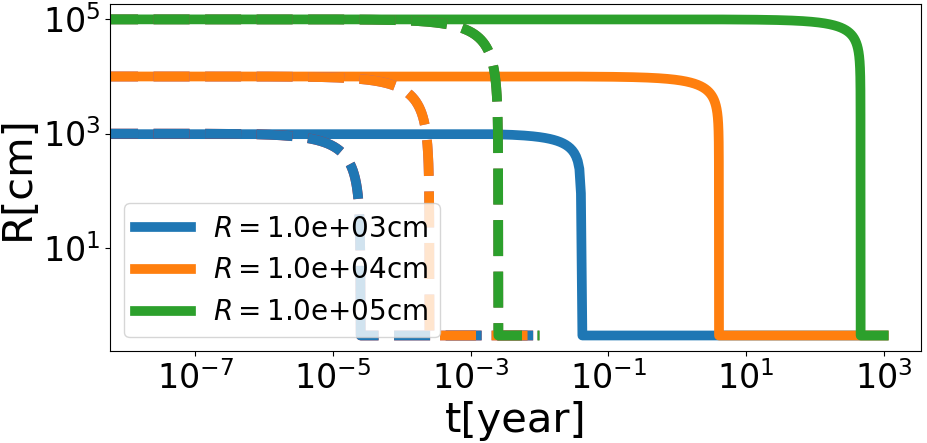}
\caption{The relationship between aeolian erosion and ablation for an icy object with initial size of $10^3 \rm{cm}$, embedded at a constant distance of $0.005 \rm{AU}$ from the WD. Solid lines describes aeolian erosion only, and dashed lines describe the combined effect of aeolian erosion and thermal ablation (above the threshold temperature for ice).  
} 
\label{fig:+ablation}
\end{figure}

As can be seen in Fig. \ref{fig:+ablation},  thermal ablation shortens the timescales in which objects are ground down to their final scales -- here manifested for icy objects. Rocky objects will require higher disk temperatures for thermal ablation to be significant, since the critical temperature for rock is $4000 \rm{K}$; due to the large range of possibilities for WD disks in general -- and for temperatures in particular -- the temperature in the disk might even exceed the critical temperature for rocks and other materials, at least for the hottest youngest WDs. For coated objects, there could be a combined process of ablating the outer icy layer and then mechanically eroding inner layers.

\section{Discussion \& Implications}\label{sec: discssion and implications}

\subsection{Parameters dependence}

The possible parameter range for WD disks is wide and enables us to study a multitude of combinations of parameters. Here we will present a parameter space exploration for solid bodies embedded in WD disks and subjected to the effects of aeolian erosion. 

 \begin{figure}
    \includegraphics[width=0.99\linewidth]{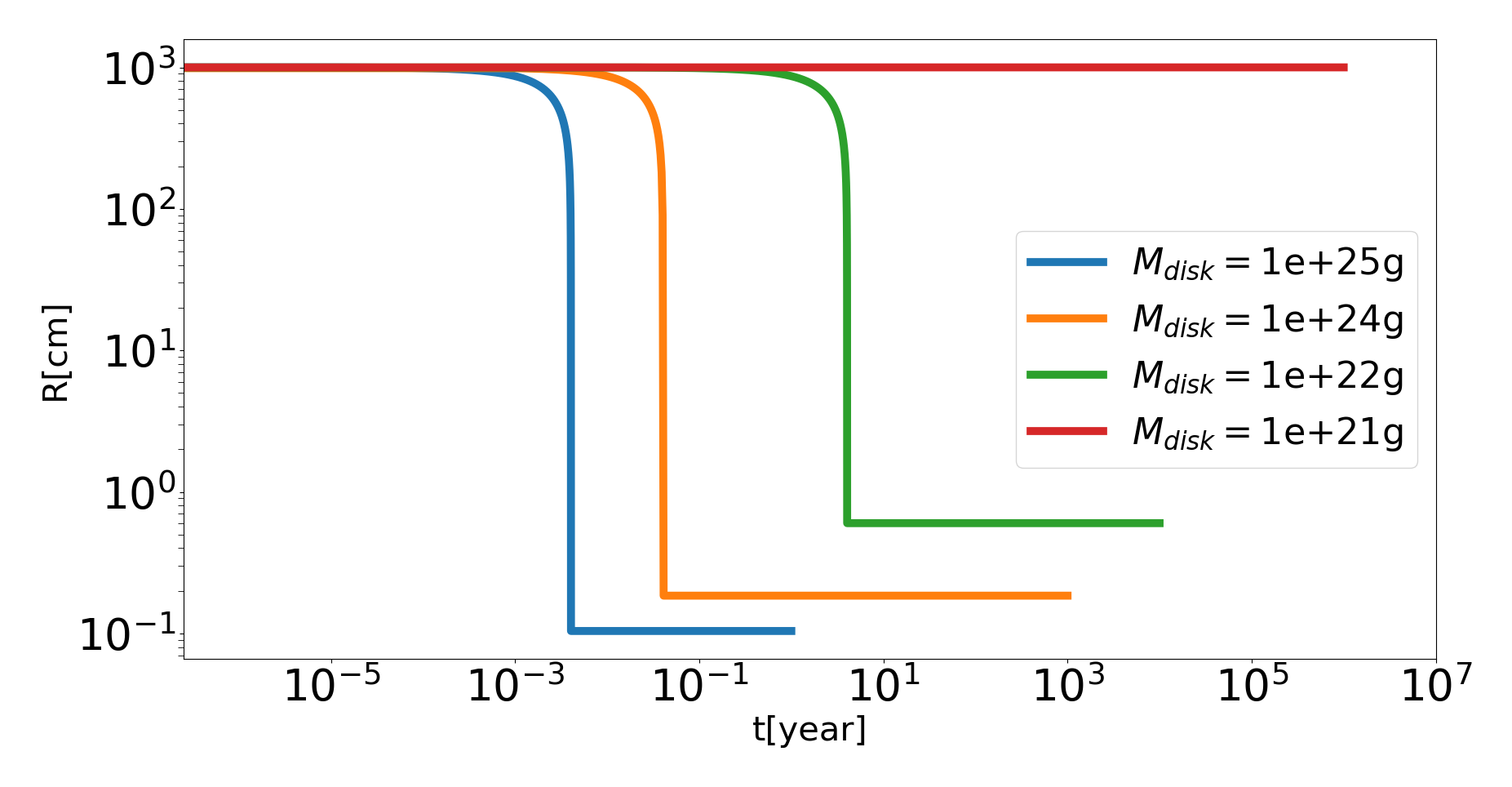}
\caption{The dependence on disk mass for an object with initial size of $10^3 \rm{cm}$, and constant distance from the center of $0.005 \rm{AU}$. 
} 
\label{fig:m_disk}
\end{figure}

In Fig. \ref{fig:m_disk} we examine the dependence of aeolian erosion on the disk mass. As can be seen in eq. \ref{eq: surface density}, the surface density of the disk, and hence the gas density, grow linearly with the disk mass. Since the the aeolian erosion rate is proportional to the gas density, the rate becomes stronger for larger disk masses. 

 \begin{figure}
    \includegraphics[width=0.99\linewidth]{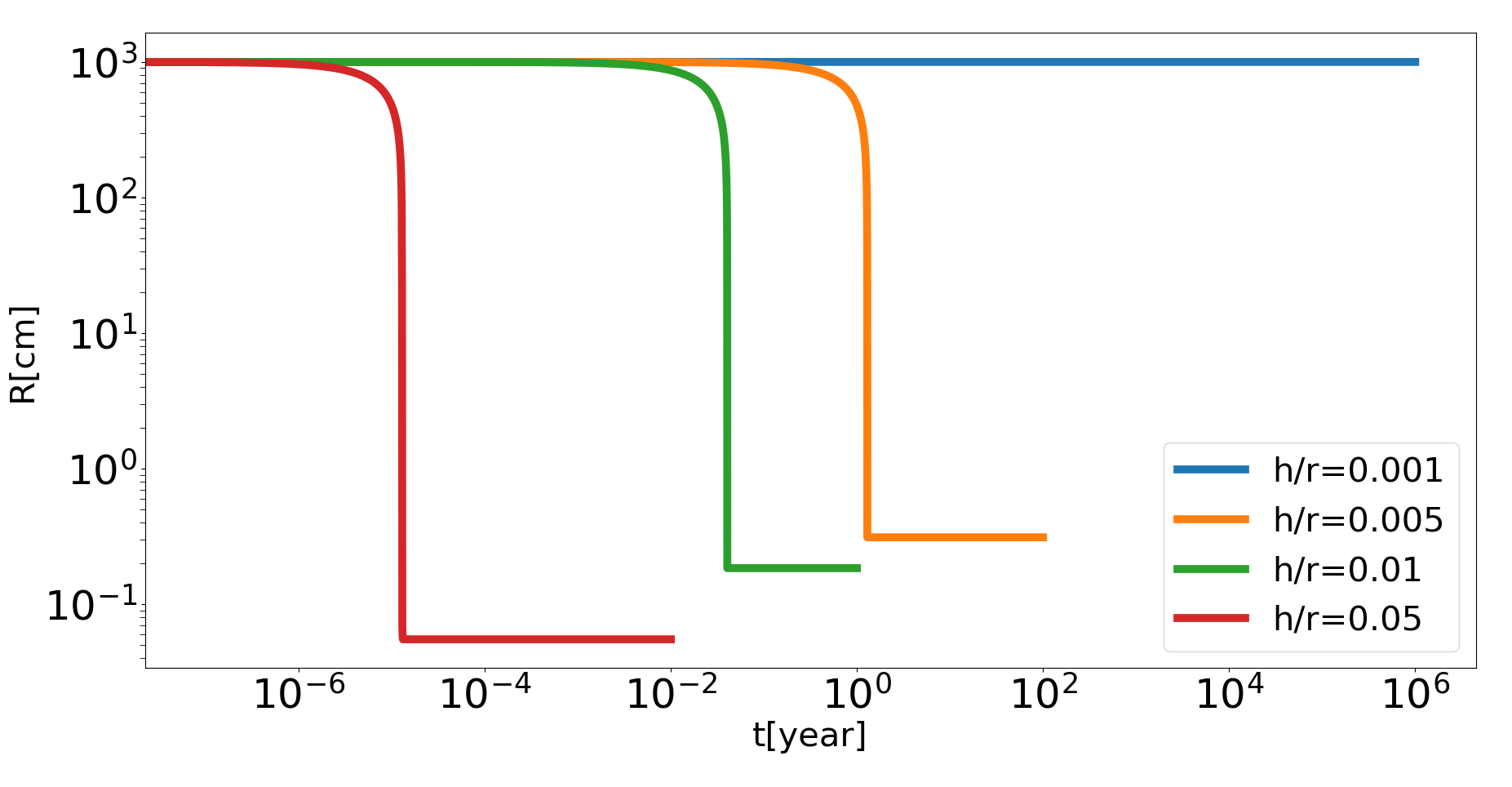}
\caption{The dependence on aeolian erosion on the aspect ratio of the disk, $h/r=c_s^2/{r\Omega}$ where $\Omega$ is the angular velocity, for objects of initial size $10^3\rm{cm}$, embedded at the disk in a distance of $0.005 \rm{AU}$.  
} 
\label{fig:aspect ratio}
\end{figure}

In Fig. \ref{fig:aspect ratio}, we present the dependence of aeolian erosion on the aspect ratio. Higher aspect ratios lead to stronger aeolian erosion. And in our default choice of parameters, the aspect ratio should be $\gtrsim 10^{-3}$ in order to maintain significant aeolian erosion. 

\subsection{Symbiotic Relations with Other Processes in The Disk}\label{subsec:symbiotic relations}
\subsubsection{Collisional Cascade}

Fragmentational collisional erosion is another destructive process that gradually diminishes the mass of objects in WD disks. This destructive process generates a collisional cascade which grinds down $10^5-10^7 \rm{cm}$ objects to $10^{-6} \rm{cm}$ objects within $10^2-10^6 \rm{years}$ (e.g. \citealp{KenyonBromley2017a,KenyonBromley2017b}). 
Collisional cascades and aeolian erosion both cause objects to lose mass, such that in the presence of gas, these two processes are symbiotic.

The aeolian erosion timescale is given by 
\begin{align}\label{eq:char_timescales}
 t_{\rm erosion}= \frac{R}{|\dot R|} = \frac{4\pi R^2 \rho_p a_{\rm coh}}{\rho_g v_{\rm rel}^3}
 .
\end{align}

\noindent
The fragmentation timescale is estimated from the collisional timescale. The collision timescale for a mono-disperse swarm is given by (\citealp{KenyonBromley_analytic2017, KenyonBromley2017a,KenyonBromley2017b} and references therein)

\begin{align}
t_0 = \frac{r_0 \rho P}{12 \pi \Sigma_g}
,
\end{align}

\noindent
where $r_0$ is the radius of all the objects in the swarm, $\rho_p$ is their density, $P$ is their orbital period and $\Sigma_0$ is the initial surface density. When a multi-disperse swarm is considered, the modification of the timescale is parametrized by a collision parameter $\alpha_c\propto (v^2/Q_D^*)^{-1}$, where $v$ is the collision velocity and $Q_D^*$ is the binding energy of the object (see \citealp{Leinhardt2012} and references therein), such that the timescale from multi-species is given by $t_c = \alpha t_0$, which can be shorter compared with the mono-disperse collision timescales. 
When the replenishment is efficient enough, the background distribution of objects remains roughly constant. However, generally, the background density changes with time as well, adding complications to the analysis. 

The ratio between the timescales of aeolian erosion and fragmentational collisions is given by 

\begin{align}\label{eq:transition cascade}
\frac{t_{{\rm erosion}}}{t_{{\rm fragment}}}=\frac{96\pi^2 h a_{{\rm coh}}}{\alpha_c Pv_{{\rm rel}}^3}R
\end{align}

\noindent
such that the timescales are comparable for a ratio around unity. From equating the ratio to unity, one can set the transition radius between the aeolian erosion dominated regime and the collisional fragmantation dominaneted regime, which depends on the rest of the parameters. 

 \begin{figure}
    \includegraphics[width=1\linewidth]{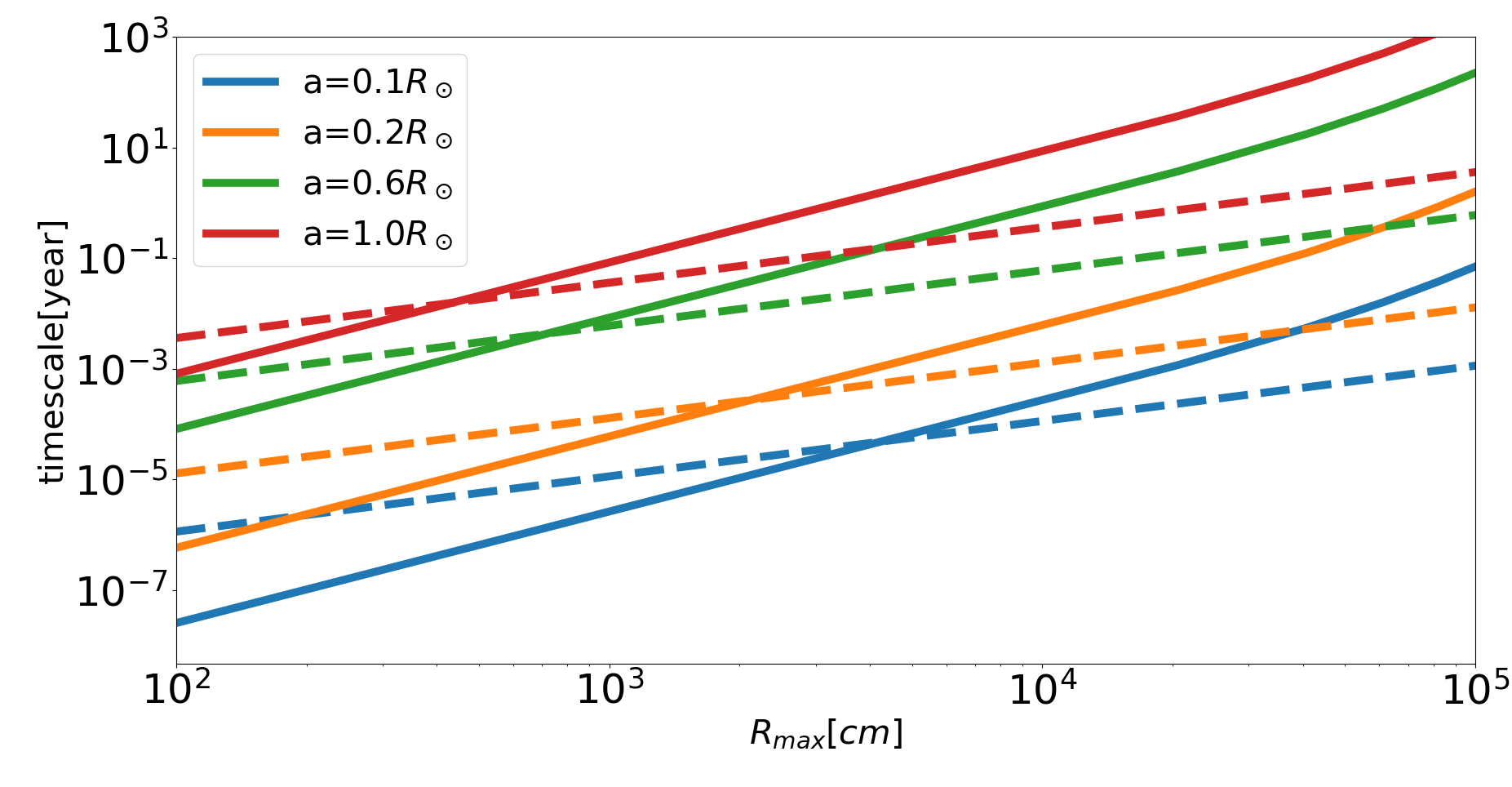}
\caption{The typical timescales of aeolian erosion (solid lines) and fragmentational collision (dashed lines) as a function of the destructed object's radius, at different constant distances from the WD.  $R_{{\rm max}}$ is the radius of the largest object in the swarm. 
} 
\label{fig:fragmentation_comparison}
\end{figure}

In Fig. \ref{fig:fragmentation_comparison}, we compare the typical timescales of aeolian erosion and fragmentational collisions at which objects are destroyed. For the fragmentation collision timescales, we consider a swarm with a maximal size given by $R_{{\rm max}}$, and assume for simplicity that the fragmentation of intermediate objects (i.e., not the largest or the smallest object in the swarm) starts just when they become the largest of the swarm, which is justified by the hierarchical character of the fragmentational cascade. For the background distribution, we assume a power law distribution of $N(r)\sim r^{-3.5}$ \citep{KenyonBromley_analytic2017}, and a value of $v^2/Q_D^*$ that yields $\alpha_c\sim 10^2$. Aeolian erosion dominates at small sizes, and fragmentational collisions at larger sizes. The transition point between the regimes varies with the distance from the WD, such that the regime of aeolian erosion grows with smaller distances. The aeolian erosion timescales are very short in these regimes, but might be comparable to or longer than the orbital period timescales of the objects. 

We would like to stress that although the destructive character of both aeolian erosion and collisional cascades leads to shrinkage of objects in the disks, they differ by their intrinsic physical mechanisms. Aeolian erosion originates from shear pressure that is induced by gas drag. Hence, the presence of gas in the disk is a necessary condition to initiate this process. The collisional cascade arises from collisions between particles in the disk.

\subsubsection{External Seeding of Objects into The Disk}
The abundance of large objects in WD disks might be replenished as a result of seeding \citep{GrishinPeretsAvni2019,GrishinVeras2019}.
Exo-planetesimals from external sources could enrich the abundance of objects in the disk that will eventually be eroded. External capture especially contributes to the number of large objects, although the total captured mass is small compared to the dust initial abundance. Furthermore, seeding could bring into the disk new materials that might eventually end as pollutants on the WD. The capture rate is dictated by the supply rate and the capture probability; both vary with the origin and the size of the captured objects. 

The captured objects change the size distribution in the disk, and might contribute to the steady-state distribution. The injection of external objects might be at a sufficient rate to cancel out the destructive processes and maintain a distribution with larger objects than expected. The revised size distribution should be derived from a full swarm simulation which includes collisional cascade, aeolian erosion, growth and seeding.

\subsubsection{Further Disk Generations}
The parameters and characteristics of the disk might vary from one generation to another. Wide-binary evolution could lead to an evolved donor that transfers mass to its companion, such that the captured material forms a disk \citep{Perets2010,PeretsKenyon2011,Schleicher2014,Lieshout2018}.
A binary main-sequence system evolves such that the more massive star sheds material which is accreted on the secondary and forms a protoplanetary disk; the mass that is lost from the system, which is much greater than the mass transferred from one star to the other, leads to expansion of the binary orbit \citep{Veras2011,KratterPerets2012,Veras2012}. Afterwards, a second generation of debris and planets form in the pre-evolved system, such that the secondary evolves off the main-sequence and sheds material to its WD companion and a protoplanetary disk formed -- similarly to the previous stages. Finally, the binary orbit expands and a third generation debris disk is formed.

Higher generation disks might give rise to different disk composition, since metallicity varies from one generation to another, such that the formed accretion disks are metal and dust rich. Higher metallicity environments are better for planet formation \citep{FischerValenti2005}, such that the formed disks are likely to have planetary and planetesimal structures.

\subsection{Possibilities for Wind Erosion of SDSS~1288+1040}

One notable potential application of aeolian erosion theory is the disk orbiting the WD SDSS J1228+1040 \citep{Manser2019}. This disk contains a planetesimal on an orbit with $a=0.73R_{\odot}$ and $e \approx 0.54$. This orbit is fully embedded in a dusty and gaseous planetary debris disc, which extends beyond $1R_{\odot}$.

Because of this compact planetesimal orbit, and because the planetesimal has been observed over at least 4,000 orbits, it cannot represent a typical rubble-pile solar system asteroid. Instead, it likely represents an iron-rich remnant core of a planet, and harbours nonzero internal tensile strength. The internal structure and size are poorly constrained: the estimated radius range is $R = 1.2-120 \ \rm{km}$.

Hence, for only the lowest end of this size range could aeolian erosion could be effectual. Figure \ref{fig:dynamical} illustrates that km-sized objects may be eroded in timescales under $\sim 10^6$ yr. However, the structures of the objects in that figure are likely to be very different than the strong and dense planetesimal around SDSS 1228+1040. Furthermore, a potentially competing effect are bodily gravitational tides.
\cite{Veras2019} show that these tides could instigate a WD to engulf objects containing just $10^{-3}M_{\oplus}$, but only for objects with sufficiently low internal viscosities.

\section{Caveats}\label{sec:caveats}
Aeolian erosion is a very effective process in WD disks, as long as the amount of gas in the disk is non-negligible. 
The current observations of WD disks with a gaseous component cannot yet well constrain several of the disk parameters, such as the disc mass and its scale height. Also, the exact origins of the gas in such disk are not well understood, with suggested origins include sublimation (e.g. \citealp{MetzgerRafikovBochkarev2012}), grain-grain collisional vaporization and sputtering. Recently, \cite{2020MNRAS.tmp.3701M} presented another channel of gas production in white dwarf disks, via interactions between an eccentric tidal stream and
a pre-existing dusty compact disk. 

Along with the destructive processes, there could be mass influx that we did not take into consideration in this paper. As manifested in \cite{KenyonBromley2016a,KenyonBromley2017b} in the context of collisional cascade, the rate of mass input might equalize the mass loss such that objects that are a-priori expected to be pulverized and will maintain their mass for long timescales. 
However, here the combined effect of fragmantational erosion and aeolian erosion might play a role, and these two processes together will lead to mass depletion of objects.  

We have neglected any asphericity in the object, which is assured due to the lack of perfect packing efficiency of its constituent grains. Furthermore, we have neglected any asphericity that develops as a result of an aeolian erosion, which acts in the direction of the headwind, and might reshape the eroded objects. 

We focus on aggregates, with a weak outer layer. Some of the objects that could be potentially affected from aeolian erosion are destroyed already in tidal shredding or collisions. The cohesion forces hold for loosly bound objects, i.e. they describe aggregates and other forms of cohesion laws should be taken into account in case of different internal physics. Once the cohesion law is dictated, the suggested prescription of aeolian erosion will be very similar to the one we sketched in this paper.

For aeolian erosion to be effective, the dominant stripping force on the weak outer layers of the aggregates would need to be erosive rather than tidal. The strength of the tidal force can vary significantly depending on physical properties; for large homogeneous rubble-piles, this value can vary by a factor of about $2$ (around $1 R_\odot$) depending on spin and fluidity, and can cause stripping on an intermittent, yearly timescale \citep{Veras2017}. Such intermittency perhaps suggests that erosive and tidal forces may act in concert in certain cases, particularly as the aggregate changes shape.

Another aspect of the physics which we did not model is the asteroid spin barrier, which refers to the minimum spin rate at which an asteroid breaks itself apart. This barrier is well-established at about $2.2 \rm{hrs}$ in the gravity-dominated regime, for spherical rubble piles larger than about $300\rm{m}$ \citep{Pravec,Hu}. However, because this spin limit is a general function of both the asphericity of the object and its internal cohesion (see the Appendix of \citealp{VerasMcDonald2020}), incorporating this limit into our modeling would not be trivial. Our aggregates change shape through time, and spin variations are not necessarily monotonic.

\section{Summary}\label{sec:summary}

The growing number of WD disks (both gaseous and non-gaseous) that have been observationally detected and characterized leave open the possibility for constraining theoretical models for the origins and evolution of such disks. However, even at the theoretical level, there are important gaps in our physical understanding of the dynamics and processes that take place in WD disks. 

In this paper we focused on the processes of aeolian erosion, which, to date, were not considered in the context of WD disks. We made use of an analytical model for aeolian erosion in WD disks, based on our studies of such processes in protoplanetary disks as presented in \cite{Erosion1}.
We find that the typical timescales of aeolian erosion in WD disks are extremely short, with aeolian erosion grinding down even km-size objects within the disk lifetime. Consequently, such processes are likely to play an important role in the evolution of small solid bodies in the disk. 
We also studied the relationship between aeolian erosion and other physical processes in WD disks and its amplification due to the combined effect (see subsections \ref{subsec:thermal destructive},\ref{subsec:symbiotic relations}).
Along with collisional cascade and thermal ablation, aeolian erosion grinds down efficiently large objects into small ones with a characteristic final size. The eroded objects experience dynamical processes that finally grind down planetesimals/rocks/pebbles/boulders into sufficiently small particles such that these could drift towards the WD via Poynting-Robertson drag and contribute to its pollution.

Aeolian erosion is the most efficient and becomes the dominant destruction process for small objects, and the critical radius for its dominance is determined by the parameters of the disk, the physical characteristics of the eroded objects, the distance from the WD and the parameters of the collisional cascade (see eq. \ref{eq:transition cascade}). 

Similarly to protoplanetary disks, aeolian erosion in WD disks induces a re-distribution of particles size, according to the distance of the particles from the WD. Hence, aeolian erosion sets constraints on the parameters of WD disks that might narrow down the current parameter space. Due to the extremely short timescales of aeolian erosion, it is not likely that observable variations in WD disks would be explained by replenishment from aeolian erosion.

\section*{Acknowledgments}
We thank the anonymous reviewer for useful comments which have improved the manuscript.
MR and HBP acknowledge support from the
Minerva center for life under extreme planetary conditions, the Lower Saxony-Israel Niedersachsisches Vorab reseach cooperation fund, and the European Union’s Horizon
2020 research and innovation program under grant agree-
ment No 865932-ERC-SNeX. DV gratefully acknowledges the support of the STFC via an Ernest Rutherford Fellowship (grant ST/P003850/1).

\section*{Data Availability}

The data that support the findings of this study are available from the
corresponding author upon reasonable request.

\bibliographystyle{mnras}

\appendix

\section{Table of commonly used parameters}

In this appendix, we present the default values for the used parameters, unless stated otherwise.

\begin{table*}
	\begin{tabular}{c| c| c| c}
		Symbol & Definition & Fiducial Value &  Reference\\
		\hline
	$\gamma$ & & $0.165 \rm{g/sec^2}$ & \cite{Kruss2019}\\
	$A_N$ & & $1.23\times 10^{-2}$ & \cite{ShaoLu2000}\\
	$\beta$ & & $ 10^{2} \rm{  g\ s^{-1}}$ & scaled from \cite{Paraskov2006} and refs. therein\\
		$\rho_p$ & planetesimals' density & rock $3.45 \rm{g} \ \rm{cm}^3$, ice $1.4 \rm{g} \ \rm{cm}^3$ & \cite{Pollack1996} \\ 

		$\mu$ & mean molecular weight & $3.85\times 10^{-24} g$ & \cite{PeretsMurrayClay2011} \\

		$\sigma$ & neutral collision cross-section & $10^{-15}\rm{cm}$ & \cite{PeretsMurrayClay2011}\\
		$T$ & temperature (optically thick) & $1000 \rm{K}$ & \\
		$M_{disk}$ & disk mass & $10^{24} \rm{g}$& \\
		$\Sigma_g$ & surface density profile (optically thick) & $5.1\times 10^3 \rm{g \ cm^2}$ &  \cite{GrishinVeras2019}\\
		$h/r$ & aspect ratio & $10^{-2}$\\
		$d$ & typical 'building-block' grain size & $0.1 \rm{cm}$\\
		%$R_s$ & sublimation radius & $10^{10} \rm{cm}$ & \cite{Rafikov2011}\\
		$T_{cr}$ & critical temperature &ice $648\rm{K}$, rock $4000 \rm{K}$
		&  \cite{Podolak1988}\\
		$L_s$ & particle specific vaporization energy (solid) & ice $2.83\times 10^{10} \rm{erg \ g^{-1}}$, rock $8.08\times 10^{10} \rm{erg \ g^{-1}}$ & \cite{DangeloPodolak2015}\\
		 & Size dist. in the disk (fragmentation only) & $N(r)\propto r^{-3.5}$ & \cite{KenyonBromley2017a}
		\end{tabular}
		\label{table:parametres_table}
\end{table*}

% Don't change these lines
\bsp	% typesetting comment
\label{lastpage}
\end{document}